\documentclass[a4paper,fleqn]{cas-dc}
\usepackage{amsmath}
\usepackage{mathtools}
\usepackage[linesnumbered,ruled]{algorithm2e}
\newcommand\Tstrut{\rule{0pt}{2.6ex}}     

\usepackage[utf8]{inputenc}
\usepackage[numbers]{natbib}
\usepackage[linesnumbered,ruled]{algorithm2e}
\DeclareMathOperator*{\argmin}{arg\,min}
\DeclareMathOperator*{\argmax}{arg\,max}
\def\tsc#1{\csdef{#1}{\textsc{\lowercase{#1}}\xspace}}
\tsc{WGM}
\tsc{QE}
\tsc{EP}
\tsc{PMS}
\tsc{BEC}
\tsc{DE}
\begin{document}
\let\WriteBookmarks\relax
\def\floatpagepagefraction{1}
\def\textpagefraction{.001}
\shorttitle{Leveraging Artificial Intelligence to Analyze the COVID-19 Distribution Pattern based on Socio-economic Determinants}
\shortauthors{M. Ghahramani et~al.}

\title [mode = title]{Leveraging Artificial Intelligence to Analyze the COVID-19 Distribution Pattern based on Socio-economic Determinants}                      
% \tnotemark[1,2]

\author[1]{Mohammadhossein Ghahramani}
\cormark[1]
\ead{sepehr.ghahramani@ucd.ie}

\author[1]{Francesco Pilla}
\ead{francesco.pilla@ucd.ie}
% \ead[URL]{www.sayahna.org}

% \ead[URL]{www.stmdocs.in}

\address[1]{Spatial Dynamics Lab, University College Dublin, Ireland}
 \cortext[cor1]{Corresponding author}

\begin{abstract}
The spatialization of socioeconomic data can be used and integrated with other sources of information to reveal valuable insights. Such data can be utilized to infer different variations, such as the dynamics of city dwellers and their spatial and temporal variability. This work focuses on such applications to explore the underlying association between socioeconomic characteristics of different geographical regions in Dublin, Ireland, and the number of confirmed COVID cases in each area. Our aim is to implement a machine learning approach to identify demographic characteristics and spatial patterns. Spatial analysis was used to describe the pattern of interest in Electoral Divisions (ED), which are the legally defined administrative areas in the Republic of Ireland for which population statistics are published from the census data. We used the most informative variables of the census data to model the number of infected people in different regions at ED level. Seven clusters detected by implementing an unsupervised neural network method. The distribution of people who have contracted the virus was studied.
\end{abstract}


\maketitle

\section{Introduction}
In March 11th, 2020, the Republic of Ireland's government launched a national action plan in response to COVID-19, a widespread lock-down in order to minimize the risk of illness. The impacts of pandemics such as the current COVID-19 should be explored extensively. To mitigate and recover from the negative repercussions, it is of paramount importance to study the effects on the social tissue in cities. It seems that various research is needed to thoroughly investigate, understand, mitigate and recover from the effect of this pandemic. Some studies have been focused on providing risk assessment frameworks based on artificial intelligence and leveraging data generated from heterogeneous sources such as disease-related data, demographic, mobility, and social media data \cite{Sannigrahi2020Examining,SILVA2021102574,GE2020102413,BERIA2021102616,SHOKOUHYAR2021102714}. The exposure risk of the pandemic in different environments has been assessed. Many researchers are exploring the dynamics of the pandemic in urban areas to mitigate effects and understand the impacts of COVID-19 on cities \cite{RUMPLER2020102469,SILVA2021102574,DAS2021102577}. In this area of research, four distinctive categories have received significant attention: environmental quality, socio-economic impacts, management and governance, and transportation and urban design \cite{Sharifi2020Urban}. As far as the socio-economic impacts are concerned, pandemics can substantially negatively affect people at the bottom of the socio-economic hierarchy, those with low education, low income, and low-status jobs. For instance, it has been discussed that the Black and Latino people's mortality rate is twice that of the Whites in the US \cite{Wade2020US}. The pandemics can also hit vulnerable groups of people in poor sanitary conditions. Moreover, various factors such as high density, inadequate access to health services and infrastructure facilities can exacerbate the situation \cite{Duffey2020Analysing,RAHMAN2020102372}. Different inequality issues can also make it difficult to maintain social distancing \cite{SUN2020102390}. Hence, it is essential to understand the existed relation between socio-economic inequalities and the pandemic. As discussed, such inequalities can threaten public health by making it difficult to enforce protective measures such as social distancing.

Artificial Intelligence technologies such as Neural Networks and Deep learning can play a significant role during a pandemic. They can be used to provide different platforms for social distance tracking \cite{AHMED2021102571,NAGRATH2021102692,GHAHRAMANI2021100058}, monitor and control the spread of COVID-19 \cite{BHATTACHARYA2021102589,ZIVKOVIC2021102669}. Such technology has been used in this study. We assess the association between the demographic features and the number of confirmed cases at Electoral Divisions (i.e., ED) in Dublin, Ireland based on an optimized self-organizing neural network. It should be mentioned that the number of cases until September 10, 2020, have been considered in this work. Our aim is to understand the impacts of the pandemic on Dublin city given associated characteristics and study the related patterns in different clusters obtaining from demographic information, i.e., census data. We used a machine learning method based on an unsupervised learning approach to group spatial data into meaningful clusters \cite{GHAHRAMANILifeInssu}. In doing so, the similarities among spatial objects were taken into account. Given the implemented model, the implicit information about different EDs were extracted, and all associated relations were examined. Such data exploration can help us extract demographic information related to various clusters. First, a feature selection method was used to extract the most relevant variables since the census data includes over 700 features, and redundant features can significantly affect the model accuracy. Feature extraction aims to project high-dimensional data sets into lower-dimensional ones in which relevant features can be preserved. These features, then, were used to distinguish patterns. Dimensionality reduction and feature selection/extraction methods \cite{Ghahramani2020feature}, e.g., principal component analysis (PCA), linear discriminant analysis (LDA), and canonical correlation analysis (CCA), play a critical role in dealing with noise and redundant features. These methods were used as a pre-processing phase of data analysis and helped us obtain better insights and robust decisions.

Broadly speaking, dimensionality reduction is considered as a method to remove redundant variables. This technique can be regarded as two distinctive approaches, i.e., feature extraction and feature selection. Feature extraction refers to those techniques that project original variables to a new latent space with lower dimensionality, while feature selection methods aim to choose a subset of variables such that a trained model minimizes redundancy and maximizes relevance to the target feature. In this work, we deal with a clustering problem and high-dimensionality issue; hence, a feature extraction technique was used. Since interpreting associated patterns in feature extraction methods can be a subjective process, different tests were implemented to deal with related issues such as readability and interpretability. PCA is a classic approach to dimensionality reduction (feature extraction) and has been implemented in various research studies. However, it suffers from a global linearity issue. Thus, to address this concern, a nonlinear technique (i.e., kernel PCA \cite{Kim2020Simple}) was used in this work.

\begin{figure}
  \includegraphics[height=10cm]{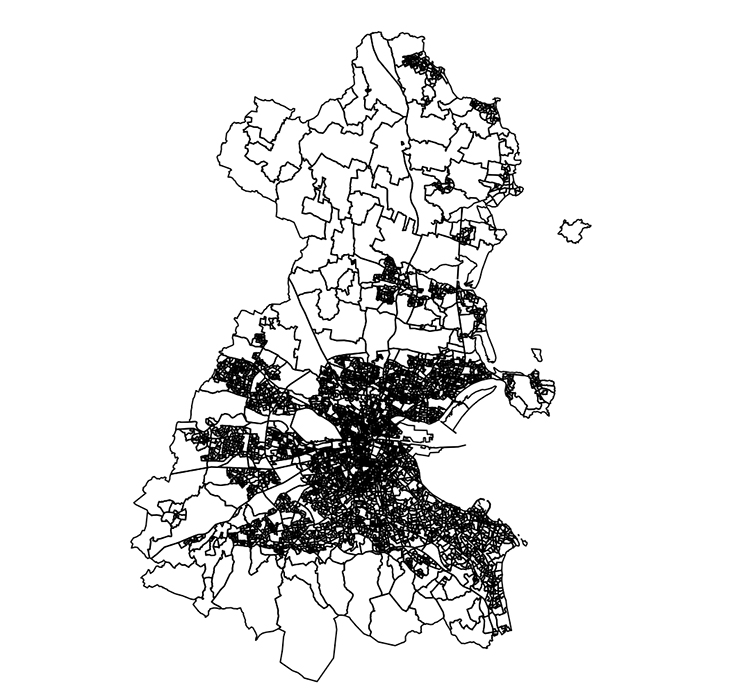}
  \caption{Dublin shapefile including different polygons of the administrative boundary and attributes of geographic features.}
  \label{ShapeFile}
\end{figure}

Then, the extracted features from the census data were fed into a clustering model, and different clusters were identified. The goal in this phase is to cluster EDs (including various demographic variables) such that similarities among them within each group are maximized. The model is based on an advanced spatial clustering technique and can deal with non-linear relationships between features of a high dimensional data set. To do so, we implemented an unsupervised approach based on an Artificial Neural Network (ANN) that can properly transform geo-referenced data into information. The main property of ANNs is their ability to learn and model nonlinear and complex relationships. The model employ a competition-based learning mechanism to generate insights from unlabelled data. It leverages a multi-layer clustering approach, i.e., a self-organizing neural network \cite{Ramos2020SOM,Yu2020Online}, to transform a complex high-dimensional input space into low dimensional output space while preserving the topology of the data. Given a set of EDs, the model groups together different spatial objects that are similar with other (i.e., the distance among observations is minimized in a given cluster). Different validity measures were also applied and the results are illustrated. For visualization, we use the shapefile of Dublin. Fig. \ref{ShapeFile} demonstrates the Dublin shapefile, including different districts.

The contributions of this work are as follows:

\begin{enumerate}

\item The link between the number of confirmed Covid cases and socio-economic determinants at Electoral Division level in Dublin, Ireland is analyzed based on an AI-based spatial clustering method.

\item A topology-preserving model is implemented to explore nonlinear relationship among Electoral Divisions given the census data to characterize
the spatial distribution of city dwellers.
\end{enumerate}

The remainder of this paper is organized as follows: some related work on application of machine learning and artificial intelligence to deal with concerns related to the pandemic is described in Section \ref{RelatedWork}; data pre-processing operations including feature extraction is explained in Section \ref{DataProcessing}; the proposed approach with its associated discussions is presented in Section \ref{model}; Section \ref{result} shows the experimental settings and the clustering results; and the future work and conclusions are presented in Section \ref{conclusion}.

\section {Related Work}\label{RelatedWork}
Due to the global spread of coronavirus, many researchers across the world are working to understand the underlying patterns of the pandemic from different perspectives. They are looking for effective ways to manage the flow of people and prevent new viral infections. As expected, numerous research has been undertaken as to medical concerns (e.g., diagnosis and treatment of the disease like lung disease, lung nodules, chronic inflammation, chronic obstructive pulmonary diseases) to ensure all required measures are in place. Different strategies, such as chest computed tomography imaging \cite{Xie2020Chest} and polymerase chain reaction \cite{Hu2020Weakly}, have been discussed for detecting and classifying COVID-19 infections. Artificial Intelligence (AI) approaches have also been used in the field of medical data analysis \cite{BHATTACHARYA2021102589}, and different algorithms have been implemented for such analysis and patients' classification. Different neural network techniques have been utilized for diagnosis based on identified clinical characteristics such as cough, fever, sputum development, and pleuritic chest pain \cite{Li2020Artificial,Ouyang2020Dual}. Various impacts of the pandemic on urban areas have also attracted the attention of researchers. In \cite{Alsaeedy2020Detecting}, the authors have introduced a novel method to identify regions with high human density and mobility, which are at risk for spreading COVID-19 by exploiting cellular-network functionalities. In doing so, they have used the frequency of handover and cell selection events to identify the density of congestion. Several visualization techniques like Class Activation Mapping (CAM) \cite{Sun2020Vision}, Class-specific Saliency Map, and Gradient-weighted Class Activation Mapping (Grad-CAM) \cite{He2020MediMLP} has been used to generate localization heatmaps in order to highlight crucial areas that are closely associated with the pandemic. Rustam et al., have implemented four Machine Learning models, such as linear regression, least absolute shrinkage, and selection operator, support vector machine, and exponential smoothing to understand the threatening factors of COVID-19 \cite{Rustam2020Future}. Different features, such as the number of newly infected cases, the number of deaths, and the number of recoveries have been taken into account in their model.

Network analysis, as a set of integrated techniques, can be used to provide direct visualization of the pandemic risk. By illustrating the degree of similarity among various areas given confirmed cases, So et al. have demonstrated that network analysis can provide a relatively simple yet powerful way to estimate the pandemic risk \cite{So2020Visualizing}. Such analysis can also supplement traditional modelling techniques to improve global control and prevention of the disease and provide more timely evidence to inform decision-making in crisis zones. In \cite{Montes2020Identification}, the authors have presented a methodology to identify spreaders using the analysis of the relationship between socio-cultural and economic characteristics with the number of infections and deaths caused by the virus in different countries. The authors have explored the effect of socioeconomics, population, gross domestic product, health, and air connections by solving a vertex separator problem in multiplex complex networks.

\begin{figure}
  \includegraphics[height=10cm]{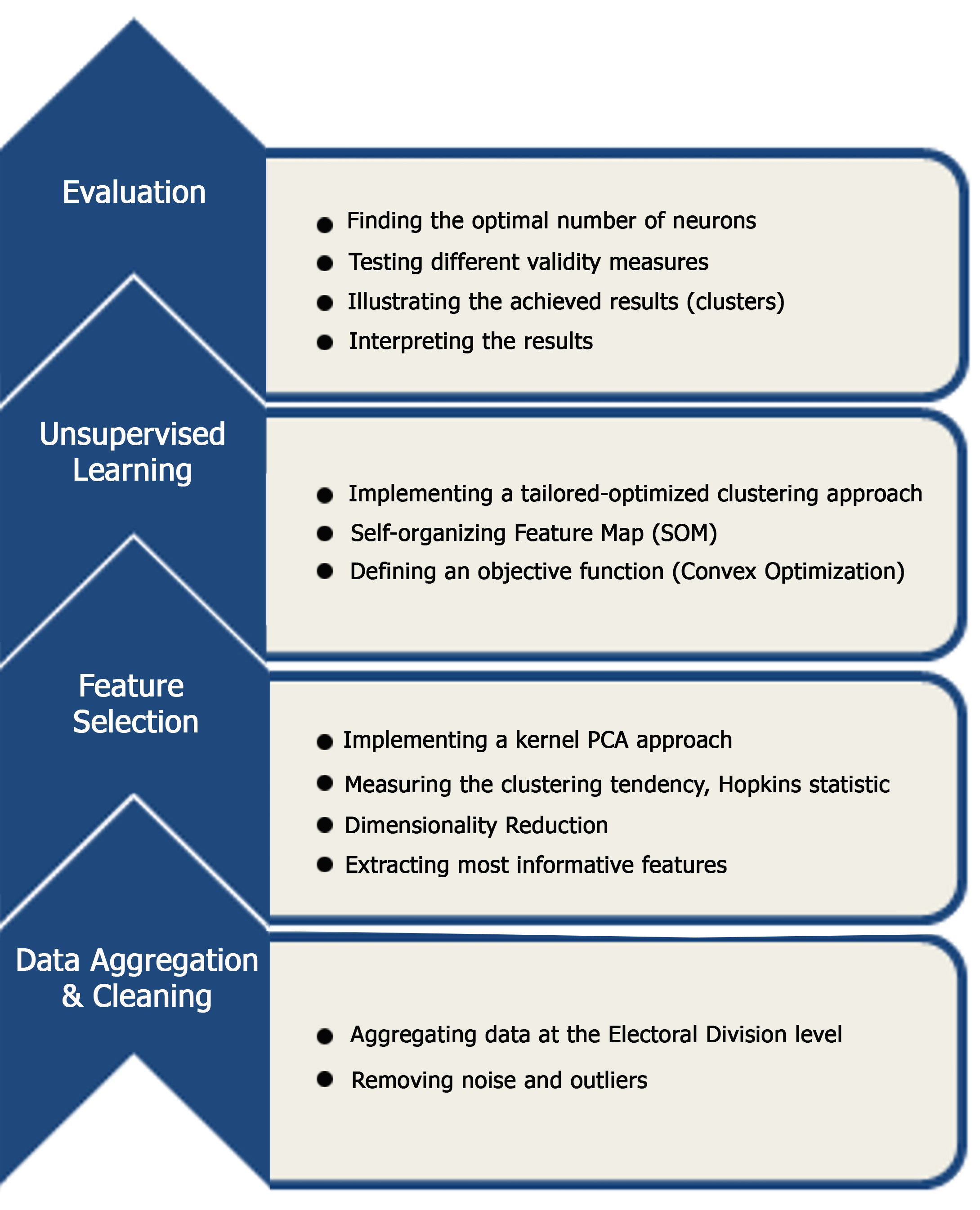}
  \caption{Different phases of the analysis model used in this work.}
  \label{diagram}
\end{figure}

Targeting policy responses to crises such as the current pandemic and interventions exclusively on people who live in deprived areas requires insights such as which clusters in society are most affected. In this work, we explore demographic and socioeconomic factors and investigate the role of socioeconomic factors in the spread of COVID-19. Our aim is to analyze underlying features obtained from census data and describe such demographic information concerning the geolocation of patients. We study the link of the pandemic with such factors. Fig. \ref{diagram} illustrates different phases of the proposed model.

\section {Data Processing}\label{DataProcessing}
Geodemographic is referred to as the study of spatial patterns and socio-economic characteristics of different areas. Associated demographic databases, such as census data, can be used to understand population diversity better since they include characteristics of a country's inhabitants. Generally speaking, Spatio-temporal datasets can be divided into different categories, such as geo-referenced data points, geo-referenced time series, moving objects, and trajectories. The estimation of a region's population has been a critical application of geospatial science in demography. In this sense, geodemographic clustering can be considered as a tool to understand spatially dependent datasets. This kind of clustering is unsupervised learning that groups spatial data into meaningful clusters based on similarities among various areas. The learning procedure is correlated to the tendency of people to associate themselves with others who have common characteristics. Census data can be considered as a reference for overall population estimation. It includes information about individuals who have been counted within households in different regions. Such data sets have some special characteristics such as geospatial features. They consist of measurements or observations taken at specific locations, referenced by latitude and longitude coordinates and/or associated within specific regions (in this work Electoral Divisions). Census data for the population living in the Republic of Ireland are available at a different level, i.e., Small Area and Electoral Division (ED), from a survey taken in 2016. However, since the number of confirmed cases are available at EDs, the census data at such administrative areas were incorporated.

\subsection {Dataset}
Demographic information is available at the local population level via censuses carried out by countries. In Ireland, a census is conducted at five-year periods by the government, with the most recent census prior to this work occurring in 2016. The census of Ireland is disseminated by the Central Statistics Office (CSO) and provides a vast amount of information. Spatial data like a census typically involves a large number of observations, meaning analysis of this nature tends to involve complex multivariate analysis and machine learning methods \cite{ghahramani1,ghahramani2,ghahramani3}. There are 322 EDs in Dublin, and the census consists of 764 features (relating to, for example, age, household size, marriage status, and education levels etc.) for each of 322 EDs. The census reports the features as a count of people. We converted these features to percentages of the population within each ED. Some sample records are presented in Table \ref{tab:records}. The number of Covid cases are also aggregated in this Table. There are no missing values or outliers in the census data. The dataset were normalized; the variables were scaled and transformed so that they each make an approximately equal contribution to the results. For example, there are about 100 variables relating to age information in the raw census data that they are summarized into percentages of different age bands; and there are about 40 variables relating to education levels that are converted to percentages of people holding a third-level higher education degree and above for each area. Take some variables demonstrated in Table \ref{tab:records} as an example. The variables T1-1AGE0M, T1-1AGE1M, T1-1AGE2M, T1-1AGE3M, and T1-1AGE4M, which refer to the number of people in different age bands (infants to four years old) have been merged, and a new feature Age0-4 has been created. In total, we extracted 53 variables that are synthesized from the census data, and a subset of these variables is presented in Table \ref{data2}. For the sake of brevity, not all summarized census variables are presented and discussed in detail. All the features created in this phase are used in a dimensionality reduction phase to be explained later. It should be mentioned that spatial features cannot be illustrated or modelled in a simple way due to their complex characteristics, e.g., size, boundaries, direction and connectivity. Hence, spatial analysis is more sophisticated than relational data processing in terms of algorithmic efficiency and the complexity of possible patterns because interrelated information at a spatial scale has to be considered. Therefore, spatial or geodemographic clustering is used for grouping and labelling geographical neighbourhoods in terms of their social and economic characteristics. Such an approach can be used to understand our spatially dependent data and the potential underlying associations between this data and confirmed number of Covid cases. Such applications allow similarities between patient structures in different EDs to be highlighted, geodemographically speaking.

\begin{table*}[ht]
\caption{Some observations of the census data at electoral divisions level consisting of $764$ variables.} % title name of the table
\centering % centering table
\footnotesize
\begin{tabular}{c c c c c c c c c c c c c c} % creating 10 columns
\hline\hline % inserting double-line
%\Tstrut
 GEOGID & GEOGDESC & T1-1AGE0M & T1-1AGE1M & T1-1AGE2M & T1-1AGE3M & T1-1AGE4M&...&T15-3-N& T15-3-NS & Covid cases
\\ [0.5ex]
\hline % inserts single-line
% Entering 1st row
% \Tstrut
E02008 & Ayrfield &33	&33	&34	&31	&37	&...&341	&43 &	133     \\[0.5ex]

\hline % inserts single-line
% Entering 1st row

%  \Tstrut
E02012	&	Ballygall B	&10	&10	&5	&8	&11	&...&266&27	&109     \\[0.5ex]
\hline % inserts single-line
% Entering 1st row
% \Tstrut
E02022	&	Beaumont B	&29	&26	&35	&24	&21 &...&270&38 & 75
    \\[0.5ex]

\hline % inserts single-line
% Entering 1st row
% \Tstrut
E02006	&	Ashtown A	&100&	84	&70	&66	&49 &...&626& 111 & 99
    \\[0.5ex]
\hline % inserts single-line
% Entering 1st row
% \Tstrut
E02093	&	Whitehall D	&11	&15	&12	&11	&5 &...&258& 16& 150
    \\[0.5ex]

% [1ex] adds vertical space
\hline % inserts single-line
\end{tabular}
\label{tab:records}
\end{table*}

\begin{table*}
\centering 
\small
\begin{tabular}{l c c c c c} 
\toprule % Top horizontal line
& \multicolumn{5}{c}{\textbf{Statistics}} \\ 
\cmidrule(l){2-6} 
\textbf{Features } & Mean  & Std deviation& Median Absolute Deviation & IQR & Median\\ % Column names row
\midrule % In-table horizontal line
Percentage of population aged 0-4 & 7.298 & 2.168 & 1.425 & [5.797, 8.638] & 7.238\\ % Content row 1
Percentage of population aged 5-14  & 14.053 & 3.379 & 1.964 & [12.272, 16.228] & 14.313\\ % Content row 2
Percentage of population aged 65 and over  & 13.580 & 4.413 & 2.620 & [10.721, 16.071] & 13.243\\ % Content row 3
Percentage of single population  & 56.157 & 4.881 & 2.432 & [53.146, 58.103] & 55.468\\ % Content row 4
Percentage of house-share household  & 4.254 & 4.147 & 1.389 & [3.112, 5.984] & 4.347\\ % Content row 5
Percentage with higher education degrees  & 20.471 & 9.131 & 4.292 & [14.908, 23.724] & 18.501\\ % Content row 5
Percentage of professional social class  & 4.981 & 3.816 & 1.863 & [2.511, 6.417] & 4.098\\ % Content row 5
Percentage of unemployed population  & 11.015 & 3.938 & 2.436 & [8.241, 13.249] & 10.526\\ % Content row 5
\midrule % In-table horizontal line
\bottomrule % Bottom horizontal line
\end{tabular}
\caption{Summary information on a subset of summarized variables from the Irish census data across all EDs} 
\label{data2}
\end{table*}

Each observation (EDs consisting of demographic information) can be defined as an \textit{m}-tuple ($m$ is the number of features).

Let matrix $X \in \mathbf{R}^{n \times m}$ as:

 \begin{equation}
X=\begin{bmatrix}
         X_{1} \\   X_{2} \\ \vdots \\ X_{n}
   \end{bmatrix}
=\begin{bmatrix}
         x_{11} & x_{12} & \cdots  & x_{1m} \\   
         x_{21} & x_{22} & \cdots  & x_{2m} \\ 
         \vdots  & \vdots  & \ddots & \vdots  \\ 
         x_{n1} & x_{n2} & \cdots  & x_{nm} 
   \end{bmatrix}
  \end{equation}
where $\mathbf{R}$ is the real number set, $X_i$ is the $i^{\text{th}}$ region and its corresponding variables (\textit{m}-tuple), and $n$ is the number of all areas. As stated earlier, we deal with high dimensionality in this work. Such datasets can pose serious challenges, such as model overfitting. The more the number of variables increases, the more the chance of overfitting.

\subsection {Dimensionality Reduction}
Dimensionality reduction is the process of eliminating redundant variables. To handle such concerns, different approaches have been considered in the literature. Generally speaking, feature extraction and feature selection techniques are applied to reduce data dimensionality. In the former approach, original features are mapped to a new feature space with lower dimensionality. The latter refers to those methods that identify and select a subset of features such that the trained model (based on the selected features) minimizes redundancy and maximizes relevance to the target feature. PCA is the most common dimensionality reduction approach; however, the transformation applied is linear. But when data follow a nonlinear structure, as in our case, approximating the model by a linear method like PCA will not perform well on the original data. Likewise, Multidimensional Scaling \cite{Saeed2020Survey} and Independent Component Analysis (ICA) \cite{Feng2020Dynamic,Shi2019Independent} suffer from the linearity issue. To address this shortcoming, nonlinear techniques such as Kernel PCA, Laplacian Eigenmaps \cite{Sun2020Dimensionality}, and Semidefinite Embedding \cite{Xiang2009Dimensionality} can be used. The two first-mentioned methods have been applied in this work. The result of the Kernel PCA is illustrated to save space. We can define the variance-covariance matrix as

\begin{equation}
S = \frac{1}{n}\sum_{i=1}^{n} (X_i - \bar X)^T(X_i - \bar X)
\end{equation}

 \begin{figure*}
  \includegraphics[width=\linewidth]{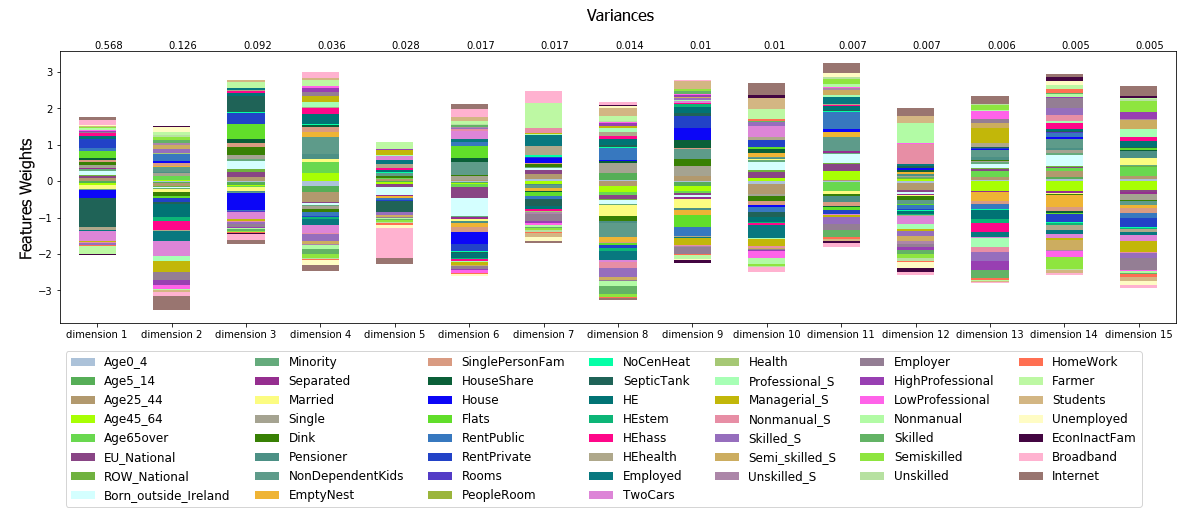}
  \caption{Result of the dimensinlity reduction phase  implemented for feature extraction based on Kernel PCA.}
  \label{kpca}
\end{figure*}

The aim is to maximize the trace of the covariance matrix (i.e., $A^*=\argmax_A tr(S)$) given a weighted covariance eigendecomposition approach \cite{Chan2012Eigendecomposition}, where $A$ is a set of eigenvectors (unitary matrices that can represent rotations of the space). A nonlinear transformation $\phi(X)$ from the original $m$-dimensional space has been considered, and the covariance matrix of the projected features has been measure as

\begin{equation}
S = \frac{1}{n}\sum_{i=1}^{n} \phi(X_i)\phi(X_i)^T
\end{equation}
The eigenvalues and eigenvectors are given by

\begin{equation}
S\nu_k=\lambda\nu_k
\end{equation}

The eigenvectors have been measured ($\nu_k=\sum_{i=1}^{n} a_{ki}\phi(X_i)$), where $k$ is the new number of dimensions.

\begin{equation}
 \frac{1}{n}\sum_{i=1}^{n} \phi(X_i)\{\phi(X_i)^T\nu_k\}=\lambda_k\nu_k
\end{equation}

By substituting $\nu_k$ in above equation

\begin{equation}
\label{eqn:1}
 \frac{1}{n}\sum_{i=1}^{n} \phi(X_i)\phi(X_i)^T \sum_{j=1}^{n}a_{ki} \phi(X_j)=\lambda_k\sum_{i=1}^{n}a_{ki} \phi(X_i)
\end{equation}
The kernel function ($\Psi(X_i,X_j)=\phi(X_i)^T\phi(X_j)$) is, then, multiply both sides of Eq. \ref{eqn:1} and the kernel principal components can be calculated as:

\begin{equation}
\phi(X)^T \nu_k=\sum_{i=1}^{n}a_{ki}\Psi(X,X_i)
\end{equation}
It should be mentioned that we have constructed the kernel matrix from the census data. To that end, a Gaussian kernel ($\Psi(X_i,X_j)=exp(-||X_i-X_j||^2/2\sigma^2)$) has been used, where $c$ is a constant. Given the measured variance for each feature, the associated weight can be measured

\begin{equation}
\sigma_X^2= \frac{\sum_{i=1}^{n}\omega_i^2(X_i-\overline{X})^2}{\sum_{i=1}^{n}\omega_i^2}
\end{equation}

We have also examined the relevance of all features using the coefficient of determination. In doing so, the proportion of the variances have been tested. A supervised learner has been used, and iteratively one feature of the dataset has been considered as the dependent variable and others as the independent variables. The Hopkins statistic, which is a way of measuring the clustering tendency of a data set, has been calculated for both scenarios with the value of $0.59$ before dimensionality reduction and $0.67$ after that phase. A value close to $1$ indicates that the data is highly clustered. Fig. \ref{kpca} illustrates the result of the dimensionality reduction given the Kernel PCA approach. Given the fraction of variances measured in this phase and also given all the weights associated to each feature, $21$ features, such as percentage of population aged 65 and over, percentage of house-share household, and percentage of the unemployed population, have been selected. All these features have been integrated with two additional variables, i.e., the population of each ED and the number of confirmed covid cases in each of those areas. The final dataset is then used in the second phase (i.e., clustering) of the model.

\section{Clustering Approach}\label{model}
After performing all the data preprocessing operations explained above, a clustering method can be implemented to find underlying patterns. Due to characteristics of this work, i.e., non-linear dynamics, an unsupervised learning mechanism based on a vector quantization technique \cite{Xie2018Adaptive} has been considered. It should be mentioned that most neural network approaches operate based on the non-linear optimization of a criterion, which may result in the local minimum issue and/or the convergence may take a long time. It has been discussed that self-organizing maps are less sensitive to such concerns. This approach is motivated by retina-cortex mapping and considered as an optimal technique for vector quantization problems. The topographic mechanism used in this method can enable us to study relationships among spatial and non-spatial features and identify associated patterns. The model is self-organized and operates based on learning rules and neuron interactions. The learning process is based on cooperation and competition among neurons. Moreover, neurons maintain proximity relationships during the learning process. The idea is to quantize the input space into a finite number of vectors. All observations in the input space (census vectors, together with the number of Covid cases in each spatial area) are projected to post-synaptic neurons in the latent space. The implemented model can transform all the census features in the input space into a low-dimensional discrete output space while preserving the relationships among variables. To do so, all vectors are mapped to neurons based on synaptic connections, each of which is assigned with weights. These weights are updated such that adjacent neurons on the lattice have similar values. The clustering procedures consists of different phases, i.e., competition, collaboration, and weight updating.

In the competition phase of the algorithm, a predefined number of neurons are initialized by randomly setting their weights using census features. Neurons compete for each input vector's ownership, and the most similar neuron (given the distance measure between an ED object together with all relevant features and all neurons) to a given observation is detected. The winning neuron is called the Best Matching Unit (BMU). There are different distance measures to find the similarity between neurons and an input vector, such as the Euclidian distance, Correlation tests, and Cosine similarity. However, the squared Euclidean distance is often used in a real application. Let $X_i$ be the $i^{\text{th}}$ input vector (i.e., $i^{\text{th}}$ ED's features) and $W_j$ the associated weights of the $j^{\text{th}}$ neuron. Then, the distance matrix $D_{ij}= \frac{1}{n} \sum_{i=1}^{n}\sum_{j=1}^{k} (X_{i}-W_{j})^2$ can be defined as:

  \begin{equation}   
  D_{ij}=
\begin{bmatrix}
    d_{11} & d_{12} & d_{13} & \dots  & d_{1k} \\
    d_{21} & d_{22} & d_{23} & \dots  & d_{2k} \\
    \vdots & \vdots & \vdots & \ddots & \vdots \\
    d_{n1} & d_{n2} & d_{n3} & \dots  & d_{nk}
\end{bmatrix}  
  \end{equation} 
The BMU can be measured according to

  \begin{equation}   
\Psi = \argmin\limits_{j} || X_i - W_j ||_2
  \end{equation}

In the collaboration phase, the adjacent neurons of a given BMU are updated. The aim is to find out which of the non-winning neurons are within the BMU's neighbourhood detected in the previous phase. To do so, the spatial location of a topological neighbourhood of the excited neuron is detected. Several neighbourhood functions can be used to calculate the neighbourhood radius, i.e., Rectangular, Mexican hat, and Gaussian functions. The latter (i.e., Gaussian function) is the most commonly used one and has been utilized in this work. The cooperative process in this phase starts with defining an initial neighbourhood radius, which shrinks throughout different iterations based on the neighbourhood function. For each neuron $j$ ($N_j$) in the neighborhood of the $i^\text{th}$ winning neuron ($N_i$), the algorithm updates all the weights associated with the $j^\text{th}$ neuron based on a learning rate. It should be mentioned that the weights of other neurons outside of $N_i$ neighbourhood are not adjusted (in a given iteration). The procedure can be defined by the function below:

  \begin{equation} 
  \label{Gaussian}  
\lambda(\xi_{ij})=exp(- \frac{\xi_{ij}^2}{2\sigma^2})
  \end{equation}
where $\lambda(\xi_{ij})$ is the topological neighborhood value of the $i^\text{th}$ winning neuron ($N_i$), $\xi_{ij}$ is a lateral distance (the distance between $\Psi_i$ and its adjacent neurons $N_j$), and $\sigma$ is a function of the number of iterations and starts with an initial value ($\sigma_o$). A decay function ($-\frac{n}{T}$) is also employed, $\sigma(n) = \sigma_o \:. \:exp(-\frac{n}{G})$, where $n$ is the number of iterations, and $G$ is a constant. By defining the distance function formulated above, the neighbourhood territory for updating all adjacent neurons is explored. Two different connections, i.e., short-range excitatory connections and long-range inhibitory interconnections, are used during the projection process. The former is utilized at the presynaptic layer and the latter at the postsynaptic one. The process can be expressed as:

  \begin{gather*}  
 \frac{\partial Y_{j}(n)}{\partial n} + \tau Y_j(n) = \\
\sum_j W_{ij}(n)X_i(n)+\sum_k \eta_{k}Y_k^\ast(n)-\sum_{k^\prime} \gamma_{k^\prime}Y_{k^\prime}^\ast(n)
  \end{gather*}
where $\tau$ is a constant, $W_{ij} (n)$ is the synaptic strength between input vectors at the presynaptic layer and neurons at the postsynaptic layer, $\eta_{k}$  and $\gamma_{k}$  are connection weights at the presynaptic and postsynaptic layers, respectively, and $Y^\ast$ is an active neuron at the postsynaptic layer.

In the third phase, two methods (i.e., Hebb's rule \cite{Wickramasinghe2019Deep,Martins2020Hybrid} and Forgetting rule \cite{Chushig2020Data}) for adjusting weights of neurons are considered. Based on the Hebb's rule, the change of the synaptic weight ($\Delta W$) is a function of relative neuron spike timing and is proportional to the correlation between an input ($X$) and an output ($Y$) of a network, i.e.,

  \begin{equation} 
  \label{Hebb}   
\Delta W = \frac{\partial W_{ij}(n)}{\partial t}=\Theta Y_j(n)X_i(n)
  \end{equation}
where $\Theta$ is the learning rate ($0<\Theta<1$). A sigmoid function has been applied during the learning process on the outputs to make sure that they are not negative.

  \begin{equation}  
Y_j(n+1) = \Phi \big[  W_j^TX(n)+ \sum_j \eta Y_j(n)  \big] 
  \end{equation}
where $\Phi$ means a sigmoid function.
Since adopting Hebbe's rule for weight updating can make weights saturated, the Forgetting rule ($\beta Y_j(n)W_{ij}(n)$) is also used in the model. Given (\ref{Hebb}) and the Gaussian neighborhood function defined by (\ref{Gaussian}), let $\Theta=\beta$, then

  \begin{equation*}   
\beta Y_j(n)=\Theta Y_j(n)=\Theta \lambda(\xi_{ij})
  \end{equation*}

we can formulate the synaptic learning rule as:

  \begin{equation}
\begin{split} 
 \frac{\partial W_{ij}(n)}{\partial t} &=\Theta Y_j(n)X_i(n) - \beta Y_j(n)W_{ij}(n) \\
&=\Theta \big[ X_i(n)-W_{ij}(n)  \big] Y_j(n)
\end{split}
  \end{equation}

With the above discussions, the weight updating process can be defined as
  \begin{equation} 
\begin{split}
W_j(n+1) & = W_j(n)+\Delta W_j \\
               & =  W_j(n)+\Theta(n)\lambda(\xi_{ij})[X(n)-W_j(n)]
\end{split}
  \end{equation}
where $\Theta(n)$ is the learning rate for the $n^\text{th}$ iteration, $W_j(t)$ is the weight vector of the $j^\text{th}$ neuron, and $\lambda$ is a neighborhood function. The learning rate is also a function of time and decreases monotonically, i.e.,

  \begin{equation*}   
\Theta(n)=\Theta_0exp(\frac{n}{-G^2})
  \end{equation*}
where $\Theta_0$ is an initial value, $G$ is a constant, and $n$ is the number of iterations.

After the weights for all the input vectors are calculated, both the learning rate and the radius are diminished. The postsynaptic weights are adjusted to resemble the census features and reflect its properties as closely as possible. To sum up the procedures, the pseudo-code of the implemented Self-organizing map is presented in Algorithm \ref{NNModel}. The summary of notations used is also given in Table \ref{table:symbols}. Two quantization and organization criteria have been utilized to measure the reliability of the model. Given such validity measures, the sensitive parameters of the algorithm have been adjusted. A discussion regarding the settings of the algorithm such as the learning rate, the size of lattice (the number of neurons), and level of similarities among neurons are presented next.

\begin{algorithm}
\SetAlgoLined

    \SetKwInOut{Input}{Input}
    \SetKwInOut{Output}{Output}
   % \Input{ $W{ij}$ Matrix (the contiguity matrix)}
    \Input{$X \gets$ Census features, $p\gets|X|$, $k\gets k_0$, $\sigma \gets \sigma_0$,
    $\Theta \gets \Theta_0$\\ $\{N_1, N_2, \cdots , N_k \} $: $k$ neurons;\\
    $\{l^{N_1} , l^{N_2} , \cdots , l^{N_k} \}$: position set;\\
    $\{w^{N_1} , w^{N_2} , \cdots , w^{N_k} \}$: initial weights;

    }
    
    \Output{neurons' weight vectors}
    {Set N = $\{N_1 , N_2 , \cdots , N_k \}$ }\;
    
        \For{$i\gets 1, \cdots, p$}{
        $ \sigma(i)=\sigma_{0}\:.\:exp(-\frac {i}{T_2})$ \\
        $ \Theta(i)=\Theta_{0}\:.\:exp(-\frac {i}{T_2})$\\
        Select the $i^{\text{th}}$ observation (ED) $x_i \in X$;\\
        $\Psi = \argmin\limits_{n \in N} || x_i - w^N ||_2$ ;\\
        \For{$j\gets 1, \cdots, k$}{
        $\xi = ||l^{x_j} - l^{\Psi} ||_2 ;$\\
        \If{$\xi < \sigma$}
    	{ $w^{N_j} = w^{N_j} + \Theta \:.\: \lambda(\xi,\sigma,n )\:.\:( x_i - w^{N_j} );$}
		}      
        }       
 
    {Output the result.}
    \caption{Pseudo-code for the SOM model}
    \label{NNModel}
    \end{algorithm}

\begin{table}[ht]
  \centering
\tiny
  \caption{Summary of the notations}
  \label{table:symbols}
\resizebox{\linewidth}{!}{
  \begin{tabular}{ll}
    \toprule
    {\bfseries Symbol} & {\bfseries Meaning} \\
    \midrule
    $X$ & Census features\\
    $p=|X|$& The number of observations\\
    $k$& Size of the lattice\\
    $\sigma$ & The neighborhood parameter\\
    $\Theta$ & The learning rate\\ 
    $\Psi$ &  The lateral distance\\
    $\xi $ & Best Matching Unit\\
    $l^{N_i}$ &Position of the $i^\text{th}$ neuron on the lattice\\

    \bottomrule
  \end{tabular} }
\end{table}

\subsection{Algorithm Convergence and Parameter Settings}
\begin{figure}
  \includegraphics[width=\linewidth]{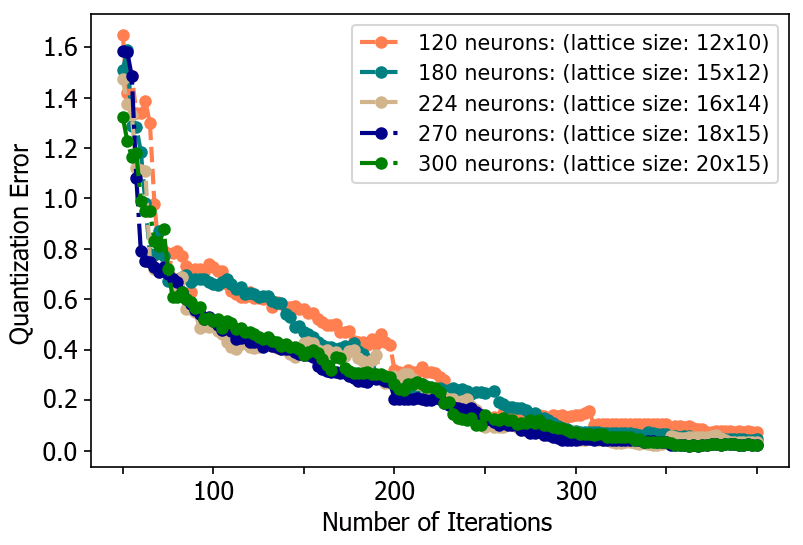}
  \caption{Comparing the Quantization Error given different lattice size}
  \label{parameters1}
\end{figure}

The learning rate and the number of units needed should be set in the algorithm, while the level of similarities among units and the proper number of clusters are designated thereafter. Different techniques can be utilized to explore the convergence of the algorithm, such as Quantisation Error (QE) \cite{Fan2018Quantization}, Topographic Error, Weight-value Convergence, and probabilistic models. It should be noted that there is no exact cost function that a self-organizing map (SOM) follows precisely. As explained before, two criteria (i.e., QE and topology preservation metric) have been taken into account to ensure that the output of the model is reliable. The quantization metric was used to assess the required number of neurons. The squared distance between an observation $X_i$ and its corresponding neuron was calculated. In other words, an optimization problem was solved based on the similarity between vectors at presynaptic and postsynaptic layers. The ultimate synaptic weights of neurons were achieved after running Algorithm \ref{NNModel}. The metric calculates the variance associated with neurons' synaptic weights by measuring the average distance between each observation and its corresponding BMU, i.e.,

  \begin{equation} 
Q_E=\frac{1}{p}\sum_{i=1}^p|| X_i - \Psi(i)||
  \end{equation} 
where $p$ is the number of observations at the presynaptic layer, summing all the errors can be expressed as:
 
   \begin{equation}
\begin{split} 
\Omega&=\sum_{i=1}^k\sum_{X^j\in V^i}\xi^2(X^j,\Psi^i) \\
&= \argmin\limits_{X^j} \xi^2(X^j,\Psi^i)
\end{split}
  \end{equation}
where $k$ is the size of the lattice (the number of neurons at the postsynaptic layer) and $V^i$ is the Voronoi areas associated with the $i^\text{th}$ BMU ($\Psi^i$). Therefore, by using such a metric for determining the convergence of the algorithm, the proper number of neurons was detected. The learning rate of the algorithm is a value between $0$ and $1$. Different initial values for the learning rate of the algorithm were tested, and the results are illustrated in Fig. \ref{parameters1}. The initial learning rate has been set to $0.57$, and $270$ neurons have been considered.

\section{Results}\label{result}

\subsection{Optimal Number of Clusters}
Given the implemented model, the algorithm leads to an organized representation of activation patterns and prototypes that well represent the census features are obtained. The next step is determining the level of similarity among neurons. We have performed different validity measures to divide neurons at the postsynaptic layer into clusters where inter-cluster similarities are minimized while the intracluster similarities are maximized. Let $C=\{C_1, C_2,...,C_m\}$ be a set $m$ clusters' centroids, $N= (N_1, N_2, \dots, N_k)$ be $k$ neurons at the postsynaptic layer and $\varphi(x_i,x_j)$ be the similarity measure between two EDs $x_i$ and $x_j$. $|N_i|^{\{m\}}$ is the number of neurons in the $m^{\text{th}}$ cluster. The first validity measure used in this work, Davies-Bouldin index (DBI), operates based on the inter-cluster and intra-cluster variance. The similarities among all ED's features projected into neurons are considered. Let denote the mean distance of all neurons belonging to cluster $C_m$ to their centroid as:

\begin{equation}\label{DBI}
    \delta_m= \frac{1}{|N|^{\{m\}}}\sum_{N_i\in Cl^{\{m\}}} ||N_i^{\{m\}}-C_m||
\end{equation}

Let $\Delta_{ij}$ be the distance between two centroids ($C_i$ and $C_j$). The Davies-Bouldin index can be formulated as: 

\begin{equation}\label{DBI}
    DBI(p)= \frac{1}{p}\sum_{i=1}^{p} max(\frac{\delta_i+\delta_j}{\Delta_{ij}})
\end{equation}
The number of clusters, i.e., $p$ in (\ref{DBI}) which minimizes the index can be considered as an optimal value.

\begin{figure}
  \includegraphics[width=\linewidth]{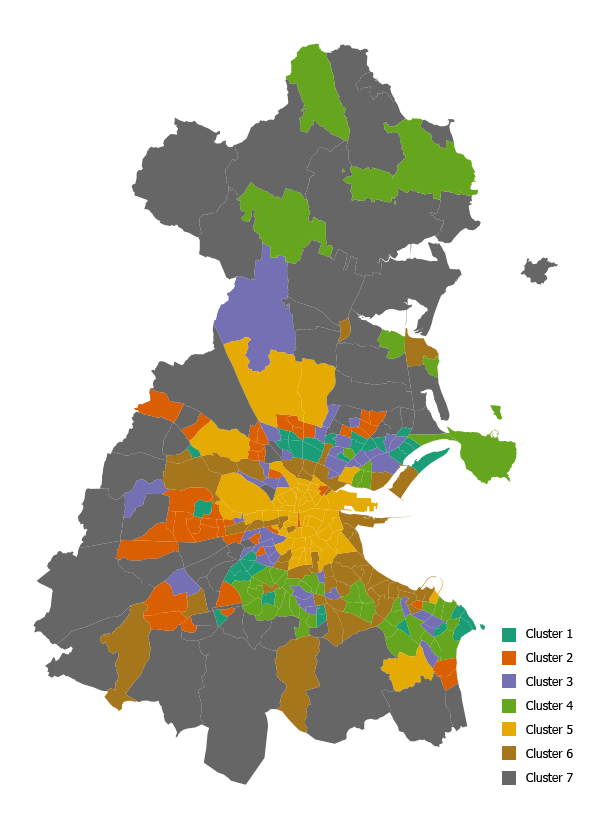}
  \caption{Clustering result of the implemented method for Electoral Divisions based on the census data, in which 7 clusters are detected; due to the fact that the small areas are dense in the city centre area.}
  \label{clusteringMap}
\end{figure}

For the second validity metric (i.e., Silhouette index), the within-cluster distance (Eq. \ref{S1}), the mean distance among neurons in each cluster ($Cl_i$), and the intra-cluster similarity (Eq. \ref{S2}) between the cluster to which $N_i$ belongs and its nearest cluster are calculated.

\begin{equation}\label{S1}
    \alpha(i)= \frac{1}{|N|^{\{m\}}-1}\sum_{N_i,N_j\in Cl^{\{m\}}} d(N_i,N_j)
\end{equation}

\begin{equation}\label{S2}
    \Lambda(N_i,C_p)= \frac{1}{|N|^{\{p\}}}\sum_{N_j\in Cl^{\{p\}}} d(N_i,N_j)
\end{equation}
The smallest intra-cluster distance is then calculated, $\beta(i)=\argmin\limits_{m\neq p} \Lambda(N_i,C_p)$. The Silhouette index ($\check{S}$) for each neuron ($N_i$) at the postsynaptic layer can be defined as

  \begin{equation} 
\check{S}=\frac{\beta(i)- \alpha(i)}{max( \alpha(i),\beta(i))}
  \end{equation} 
The mean of the index defined above for a given cluster is then calculated. Silhouette values fall between $-1$ and $1$, and a value close to $1$ indicates that the corresponding number of clusters is optimal. Considering the DBI measure, the average distance among clusters should be minimized. Hence, the minimum values for this validity index are considered. According to the results achieved from the validity measures presented in Table \ref{optimalNumberClu}, we choose seven as the optimal number of clusters. The results achieved in this work show that the algorithm converges appropriately, and the generated neural network units have been decently grouped into super-clusters. Finally, the results of the clustering method are illustrated in Fig. \ref{clusteringMap}.

\begin{table}[ht]
\caption{Two validity measures tested for selecting an appropriate number of clusters.} % title name of the table
\centering % centering table
\scriptsize
\begin{tabular}{c c c c c c c c} % creating 10 columns
\hline\hline % inserting double-line
% \Tstrut
Number of clusters & Silhouette index & Davies-Bouldin index  
\\ [0.5ex]

\hline % inserts single-line
% Entering 1st row
% \Tstrut
3 & 0.4212  & 0.1721    \\[0.5ex]

\hline % inserts single-line
% Entering 1st row
% \Tstrut
4 & 0.4961   & 0.1281   \\[0.5ex]

\hline % inserts single-line
% Entering 1st row
% \Tstrut
5 & 0.5007  & 0.0998  \\[0.5ex]

\hline % inserts single-line
% Entering 1st row
\Tstrut
6 & 0.6741  & 0.0954  \\[0.5ex]

\hline % inserts single-line
% Entering 1st row
% \Tstrut
7 & 0.8311  & 0.0704  \\[0.5ex]

\hline % inserts single-line
% Entering 1st row
% \Tstrut
8 & 0.8019  & 0.0731  \\[0.5ex]

\hline % inserts single-line
% Entering 1st row
% \Tstrut
9 & 0.7702  & 0.0782  \\[0.5ex]

% [1ex] adds vertical space
\hline % inserts single-line
\end{tabular}
\label{optimalNumberClu}
\end{table}

\begin{figure*}
  \includegraphics[width=\linewidth]{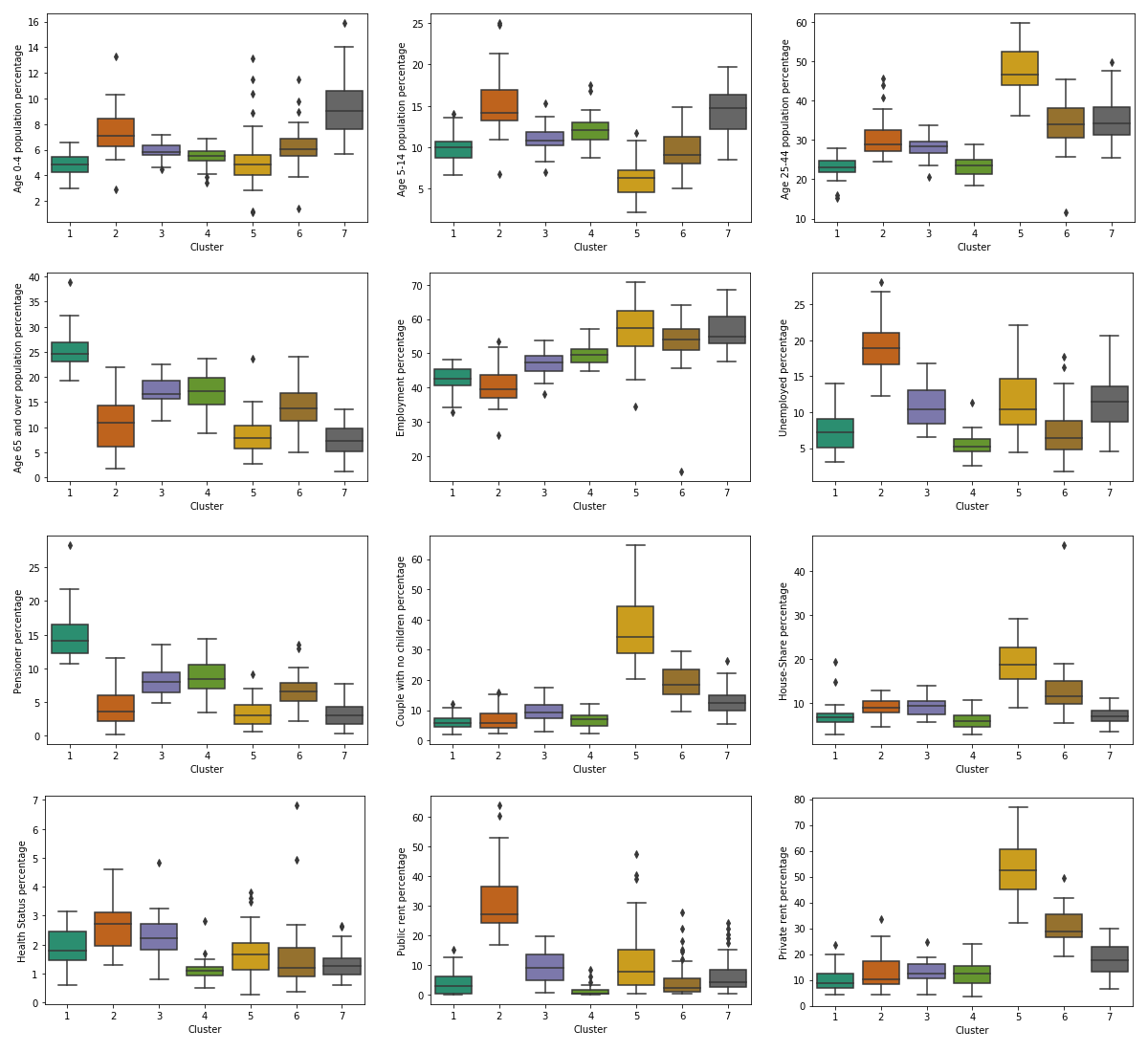}
  \caption{Boxplots of census data on percentage of different variables given 7 detected clusters.}
  \label{BOXPlot1}
\end{figure*}

We have aggregated the number of confirmed COVID cases in each Electoral Division given the identified clusters, and the results are demonstrated in Table \ref{tab:summary1}. As shown, the number of confirmed COVID cases in Clusters 5, 6, and 7 are higher comparing with others.
Given the result of the clustering model and the visualizations in Fig. \ref{clusteringMap}, we can identify different characteristics of each cluster. The detailed features are presented in Table \ref{tab:census-key}. We have found that those clusters with a high number of cases have the lowest proportions of the population with age over 65, high percentage of employment, high percentage of private rent, and high percentage of the population aged 25-44 (young professionals). At the same time, they have the highest proportion of house shares. The boxplots illustrated in Fig. \ref{BOXPlot1} correspond to the cluster characteristics in the seven detected clusters.

\begin{table}
\centering 
\small
\begin{tabular}{l c c c} 
\toprule % Top horizontal line
\textbf{Clusters } & Number of cases & Population & Cases/Pop\\ % Column names row
\midrule % In-table horizontal line
Cluster 1 &  788 &97,014& 0.0081  \\ % Content row 1
Cluster 2 & 1034&157,018&0.0065\\ % Content row 2
Cluster 3  & 901&129,784&0.0069 \\ % Content row 3
Cluster 4 & 1077&180,540&0.0059 \\ % Content row 4
Cluster 5 & 2540&271,128&0.0093 \\ % Content row 5
Cluster 6  & 1824 &171,103&0.0106\\ % Content row 5
Cluster 7  &  3635&350,772&0.0103\\ % Content row 5
\midrule % In-table horizontal line
\bottomrule % Bottom horizontal line
\end{tabular}
\caption{The number of confirmed Covid cases across seven clusters; the corresponding values of the cases/population metric for clusters 5, 6, and 7 are higher than those of others.} 
\label{tab:summary1}
\end{table}

\begin{table*}
\centering 
\small
\begin{tabular}{l c} 
\toprule % Top horizontal line
\textbf{Clusters } &Some characteristics of three clusters with high number of cases\\ % Column names row
\midrule % In-table horizontal line
Cluster 5  & $\bullet$ High percentage of house share $\bullet$ High number of couples with no child $\bullet$ High proportion of aged 25-44 \\ % Content row 5
Cluster 6  & $\bullet$ High percentage of house share $\bullet$ High proportion of dink family $\bullet$ High employment rate \\ % Content row 5
Cluster 7  & $\bullet$ High percentage of house share $\bullet$ High employment rate $\bullet$ High proportion of aged 0-14  \\ % Content row 5
\midrule % In-table horizontal line
\bottomrule % Bottom horizontal line
\end{tabular}
\caption{Some characteristics of clusters} 
\label{tab:census-key}
\end{table*}

\section{Conclusions and Future Work}\label{conclusion}
In this work, we have proposed a multiple-level approach to study the association between geodemographic clustering and the number of confirmed Covid cases in Dublin, Ireland. This work suggests that by incorporating and clustering the publicly available census data, we can obtain valuable insights regarding the spatial variations of people who have contracted the virus. The proposed method includes various phases. As the census data used in this work consists of numerous features, and such characteristics can make a predictive modelling task challenging, a feature selection approach has been implemented based on a non-linear method. Different tests have also been applied to make sure the most relevant features are selected. Then, an advanced geodemographic clustering algorithm was implemented based on a self-organizing feature map to extract clusters given the selected features. The quality of the generated map was analyzed. It should be noted that there is no universal definition of what is good clustering, and this notion is relative. As discussed throughout the paper, an SOM was considered in this work due to the inherent non-linear characteristics of the spatial dataset. Different validity measures were employed to make sure the results of the method used are reliable. We demonstrated that the algorithm has converged properly.

According to the analysis, we have detected seven clusters based on the census data and the spatial distribution of the people were explored using the unsupervised neural network method. The distribution of people who have contracted the virus was studied. The use of the proposed geodemographic approach incorporating spatial data of a geodemographic nature means that clusters can be interpreted in terms of real-life infected people attributes.

\bibliographystyle{cas-model2-names}

\bibliography{cas-refs}


\end{document}